# North Atlantic Right Whales Up-call Detection Using Multimodel Deep Learning

Ali K Ibrahim[1,2], Hanqi Zhuang[2], Laurent M. Chérubin[1], , Nurgun Erdol[2], Gregory O Corry-Crowe[1] and Ali Muhamed Ali[2]

[1]*Harbor Branch Oceanographic Institute, Florida Atlantic University, 5600 US1 North, Fort Pierce, Florida 34946, USA*

[2]*Department Computer & Electrical Engineering and Computer Science Florida AtlanticUniversity Boca Raton, FL 33431*

*Aibrahim2014@fau.edu* ,

*zhuang@fau.edu* ,

*lcherbin@fau.edu* ,

*amuhamedali2014@fau.edu* ,

*erdol@fau.edu,*

*gocorryc@fau.edu*

---

[1] aibrahim2014@fau.edu

**Abstract:**

A new method for North Atlantic Right Whales (NARW) up-call detection using Multimodel Deep Learning (MMDL) is presented in this paper. In this approach, signals from passive acoustic sensors are first converted to spectrogram and scalogram images, which are time-frequency representations of the signals. These images are in turn used to train an MMDL detector, consisting of Convolutional Neural Networks (CNNs) and Stacked Auto Encoders (SAEs). Our experimental studies revealed that CNNs work better with spectrograms and SAEs with scalograms. Therefore in our experimental design, the CNNs are trained by using spectrogram images, and the SAEs are trained by using scalogram images. A fusion mechanism is used to fuse the results from individual neural networks. In this paper, the results obtained from the MMDL detector are compared with those obtained from conventional machine learning algorithms trained with handcraft features. It is shown that the performance of the MMDL detector is significantly better than those of the representative conventional machine learning methods in terms of up-call detection rate, non-up-call detection rate, and false alarm rate.

## I. INTRODUCTION

The North Atlantic Right Whale (*Eubalaena glacialis*, NARW) is one of the most endangered species of whales in the world (Cooke, 2018). Low reproductive rates, small population size (The current population estimate for NARWs of the east coast of North America is 451 (Hayes et al., 2017), and a decreasing trend, combined with high levels of human activities (including shipping and fisheries) in their preferred habitat, underscore their precarious situation. The ability to efficiently track their numbers, migration paths and habitat use is vital toto preventing accidental injury and deaths and to promoting their recovery. Passive acoustic monitoring has proven very effective in detecting NARWs. Acoustic monitoring of NARWs can be accomplished by detecting their up-calls, their signature vocalizations, which are narrowband signals with frequency swings in the range of 50-250 Hz (Clark, 1982).

Early attempts at detection of NARW up-calls consisted of single-stage algorithms that used edge detection (Gillespie, 2004) and time-frequency domain convolution (Mellinger and Clark, 1993). They had relatively high levels of false positive errors (Urazghildiiev et al., 2009). Later methods with feature extraction and classification capabilities (Mellinger, 2004; Urazghildiiev and Clark, 2006) were able to minimize the probability of false alarm and achieved the detection probability of 0.8 or higher. The detection procedure reported by Gillespie has the following two stages. In the first stage, parameters were extracted from contours obtained from smoothed spectrograms using an edge detection method. In the second stage, these parameters were fed into a classifier to find the sounds associated with right whales. The two-stage approach was also utilized by Urazghildiiev et. Al (Urazghildiiev and Clark, 2006), who used a generalized likelihood ratio test (GLRT) detector of polynomial-phase signals with unknown amplitude and polynomial coefficients observed in the presence of locally stationary Gaussian noise.

Recently, a number of detection schemes based on neural networks have been proposed (Pylypenko, 2015). In our earlier work (Esfahanian et al., 2015), we treated NARW vocalization spectrograms as images and extracted space-time features for classification. We tested time-frequency parameters extracted from up-call spectrogram contours, and texture data output of a Linear Binary Pattern (LBP) operator applied to spectrograms. We reported a detection accuracy of 93% using these two feature sets with some popular classifiers. The use of image processing tools to extract features from NARW vocalization spectrograms was inspired by the fact that a biologist does the same task manually with direct observation of the spectrograms. Although the published method achieved a high level of detection accuracy, it required $O(N/(1-p))$ data points for a call recording of length N arranged in frames with overlap fraction $0.5 < p < 1$. One of the most popular features for classification of vocalization signals (including those of marine mammals) is the set of Mel Frequency Cepstral Coefficients (MFCCs) (De Veth et al., 2001; Mermelstein, 1976). Derived from the human auditory perception, the MFCCs are short-term spectral features (Mermelstein, 1976) of human speech and are often used for speech recognition. They are, however, not very effective under noisy conditions(Mermelstein, 1976) . Mel-frequencies are empirically derived to match the human auditory perceptional sensitivities and are logarithmically spaced. For non-human vocalizations, there is no known reason, other than convenience, why the cepstral coefficients should be derived from the Mel-frequency spectra as opposed to any log spectra. In our earlier work, wavelet techniques were also introduced for the purpose of filtering ambient ocean noise. The structures of the wavelet filters and kernel transformations establishing hyperplane decision boundaries were investigated to improve the effectiveness of MFCCs for NARW up-call detection.

Recent studies also demonstrated that deep learning-based detectors and classifiers might not need sophisticated preprocessing and handcrafted feature extraction procedures. Deep learning algorithms, such as autoencoders, convolutional neural networks (CNNs), and recurrent neural networks (RNNs), can act as feature extractors and classifiers. CNNs are especially effective in identifying spatial patterns from images. On the other hand, RNNs are known to be capable of extracting discriminative patterns from time signals. However, the phenomenon of vanishing gradients prevents a standard RNN from memorizing long-term dependency of an input time sequence. Long short-term memory (LSTM) networks solve this problem by introducing parameters that selectively memorize or forget certain attributes of an input sequence.

In this study, we use a MMDL detector, consisting of a number of Convolutional Neural Networks (CNNs) and Stacked AutoEncoders (SAEs), to detect NARW up-calls from spectrograms. In this scheme, a set of CNNs and SAEs are randomly generated to form an ensemble of classifiers. The outputs of these models are then used to train a fusion mechanism, and to determine the class of a given input once the model is trained. Through this research, we hope to provide a computational tool that is capable of accurately detecting NARW up-calls. The remainder of the paper is organized as follows. Section 2 outlines feature extraction methods, including both the MFCC and the GFCC algorithms. Section 3 discusses the detection method. Section 4 presents detection results with a database of NARW up-calls. Concluding remarks are given in the final section.

## II.   NORTH ATLANTIC RIGHT WHALE SOUNDS

Right whales vocalize a variety of low frequency sounds, and the calling repertoires between the three species are considered to be similar. So-called moans, groans, belches, and pulses have most of their acoustic energy below 500 Hz. Some vocalizations will occasionally reach up to 4 kHz. One typical right whale vocalization used to communicate with other right whales is the "up-call". It is a short "whoop" sound that rises from about 50 Hz to 440 Hz and lasts about 2 seconds. Up-calls are often described as "contact" calls since they appear to function as signals that bring whales together (ref.). Figure 1 shows different up-call types, some of which have one chrip, and others have more than one chrip. Moreover, passive acoustic data typically contains different types of noise and other sounds. These sounds are shown in Figure 2.

## III.   MULTIMODEL DEEP LEARNING

CNNs have proven to be one of the most effective deep learning structures for image classification and identification (Gu et al., 2018). CNN networks became popular after an exceptional performance of AlexNet in the 2012 ImageNet competition (Krizhevsky et al., 2012). The basis of a CNN is similar to a simple feed forward network (FFN) in that inputs are fed through an input layer and processed through several hidden layers before outputting a final result. However, A CNN block has several different layers, including input, convolution, pooling, and activation layers, which allow them to capture representative features from the input data. The first step of a CNN algorithm utilizes sliding windows of identical weights to process inputs, allowing them to synthesize neighboring features from the input data. Afterward, a nonlinear activation function is applied to remove negative values. The result is then downsampled with a pooling layer to produce more compact features. This process is repeated many times until sufficiently discriminate and concise features are obtained. These are then fed to a fully connected layer where the final decision is reached.

An SAE is composed of a multi-layer of unsupervised autoencoder networks and a layer of fully connected with SoftMax as an activation function. The learning process of an SAE has two steps: unsupervised learning and supervised learning. First, unlabeled samples are used for autoencoder's greedy layer-wise training, in which raw data is used to feed the first layer of the SAE for unsupervised training, and then the parameter vector, $w^{(1)}$, of the first hidden layer is obtained. In each step, the hidden output of the (k–1)th trained AE layer used as input to train the kth AE layer to obtain the parameter vector $w^{(k)}$. Second, a fully connected layer with labeled data is carried out for supervised learning.

For this study, we propose an MMDL method for NARW up-call detection. The model consists of multiple models of CNNs and SAEs to identify NARW up-calls. CNNs are used in the detector because of their capability of extracting both low level and high level features from input images. SAEs are used because SAEs are naturally good at data compression, which is important for feature extraction. The output of each CNN or SAE model is piped into a fusion block to reach the final decision. The fusion block inspects the outputs from individual models in search for locally consistent, discriminate and representative patterns. Our experiments showed us that spectrograms (a visual representation of the sound) worked better with CNNs and on the other hand, scalograms (a visual representation of transformed/processed sound) worked better with SAEs. Therefore, in the MMDL system, we pair CNNs with spectrograms and SAEs with scalograms. That is, the CNNs in the MMDL detector are trained with spectrogram images and the SAEs are trained with scalogram images. For every SAE and CNN, we define a range in which hyperparameters are randomly generated. This new model can be both computationally inexpensive and structurally simpler than a deep CNN. In addition, the proposed structure creates a generalized architecture that is capable of adapting to data size, making it less likely to overfit. The architecture of the proposed MMDL detector is shown in Figure 3.

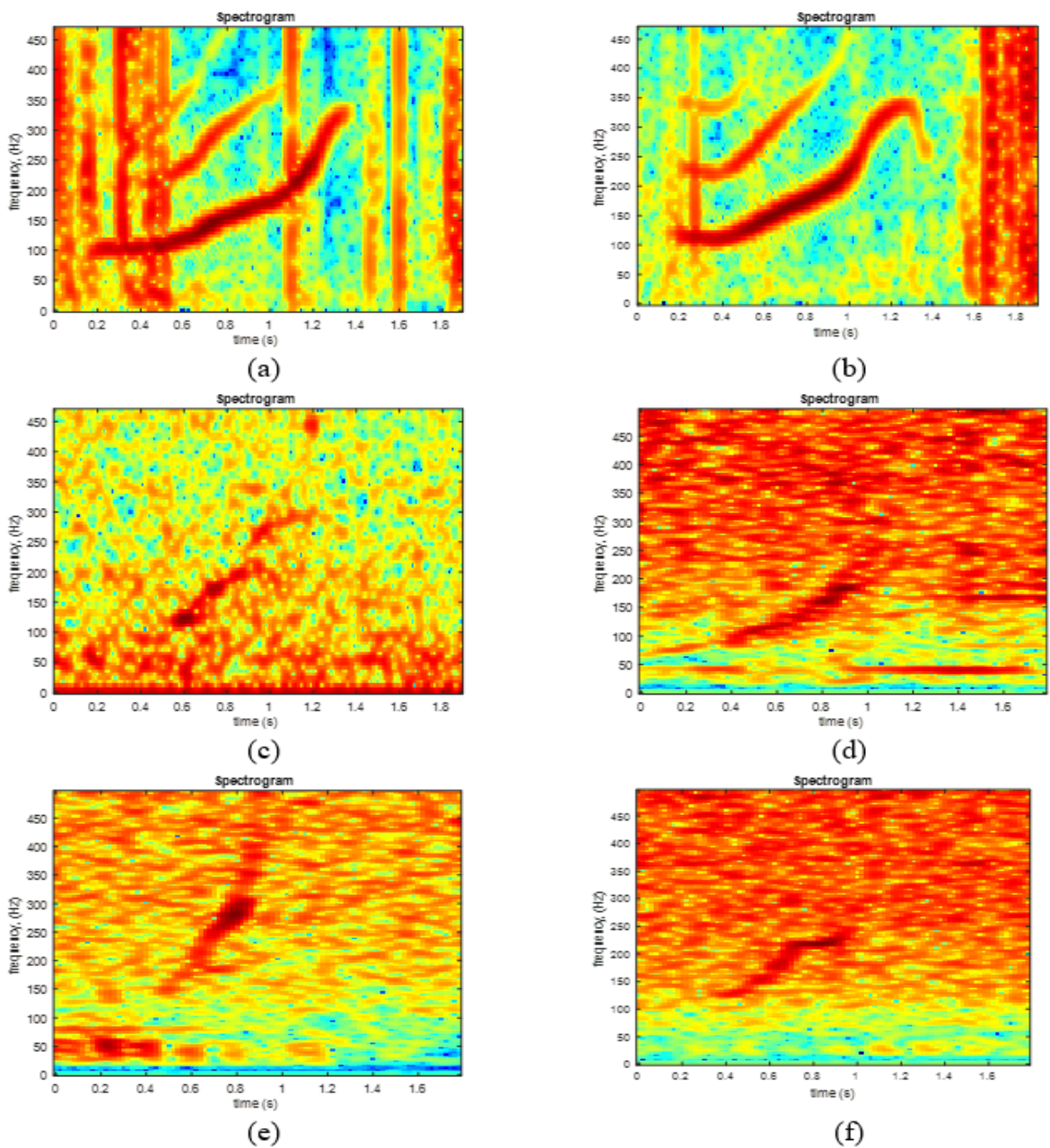

Figure 1. NARW up-call sound spectrograms

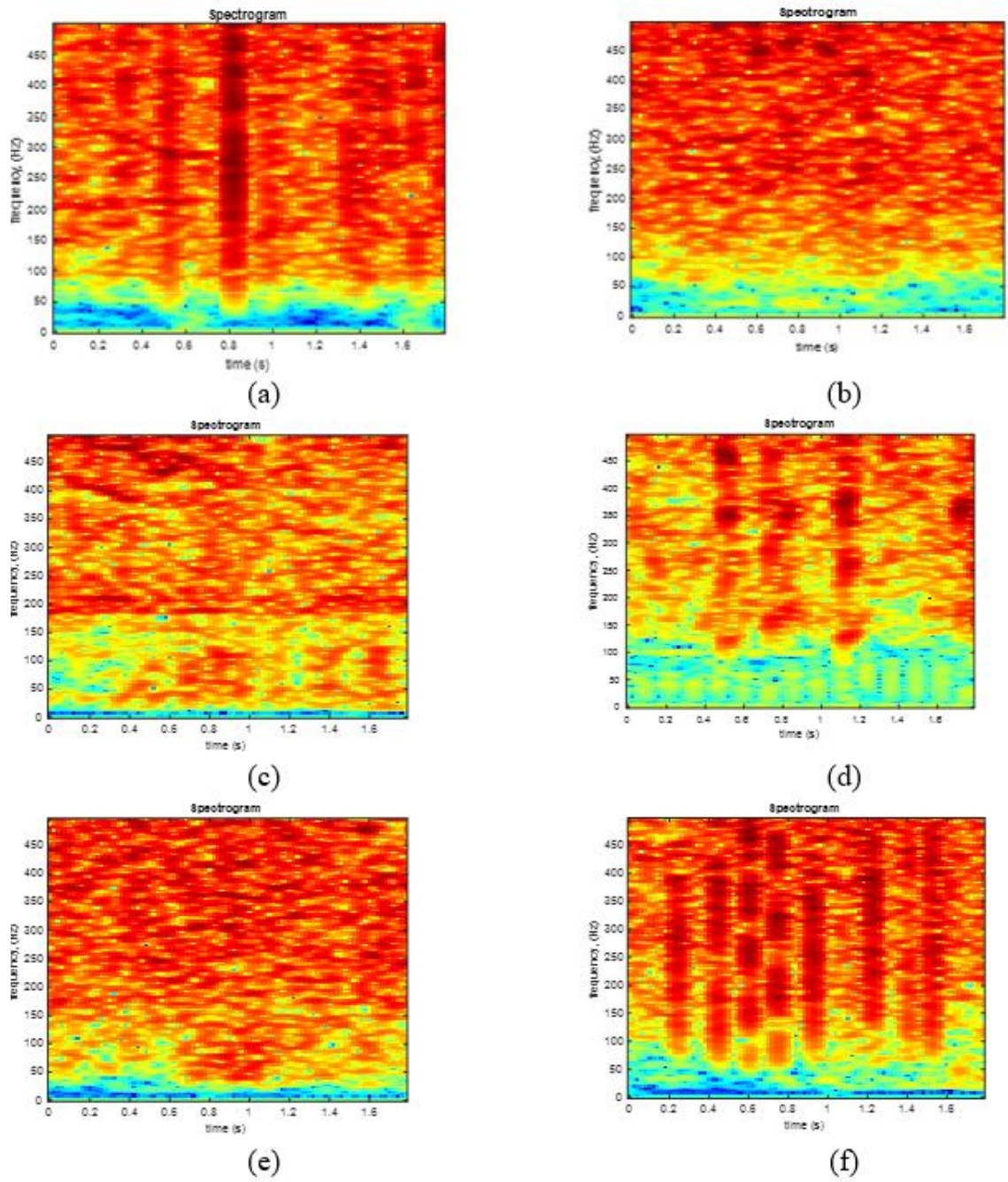

Figure 2. Background noise and other sound spectrograms

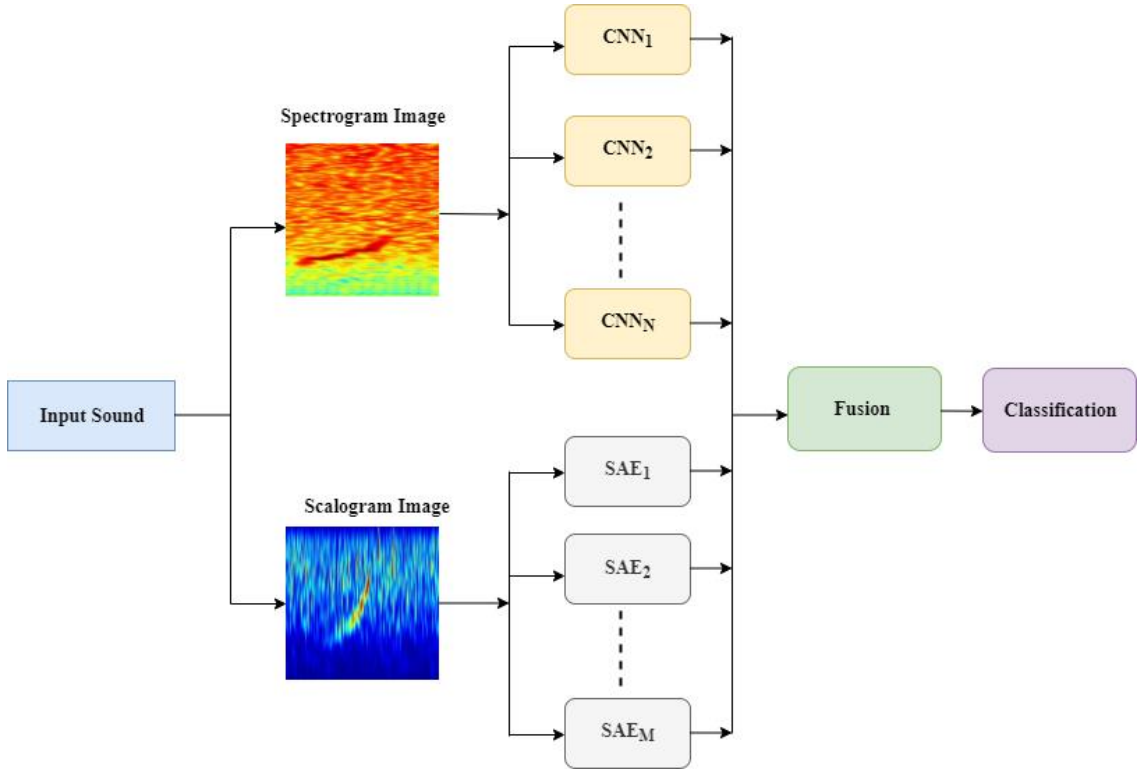

Figure 3. Proposed MMDL detector for NARW up-call detection.

## IV. IMPLEMENTATION

### A. Data Preparation

The dataset, a collection of NARW sounds, is first converted to spectrogram and scalogram images with 100x100 pixels. Figure 4 shows an example of scalogram and spectrogram images from the dataset. We used data augmentation to increase the number of training samples, and hence to make the MMDL detector more robust. The data was augmented by applying affine transformations and injecting noise to the original data samples so that our network can learn to be insensitive to orientation variations and noise in the real world. After applying random rotations, scal-ings, reflections, and shearings, the dataset contained approximately 5000 images per class as compared to 2000 images per class prior to data augmentation. We train our network structure on the augmented dataset.

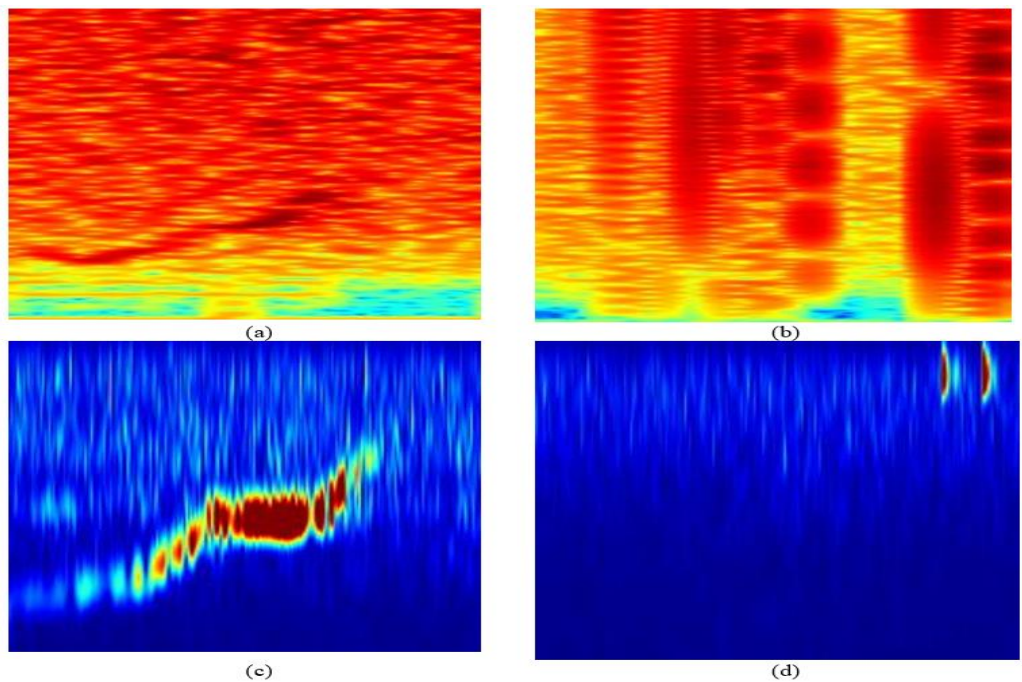

Figure 4. Spectrograms and scalograms of the datasets.

**B. System Overview**

Usually, a neural network-based machine learning algorithm requires fine-tuning to work with a specific dataset. While the approach has the advantage of avoiding hand-crafted features, the proposed MMDL method further automates the architecture construction process. By utilizing a wide range of hyperparameters and randomization, our system effectively learns the best network architecture. Furthermore, our methodology does not simply default to the best "chosen" model, instead, we use the "opinions" of individual models through a fusion mechanism. This ensures that our system takes full advantage of the strengths of both good models and outliers. The algorithmic steps of training the MMDL detector are given as follows.

1. Prepare a training dataset by converting sound files to spectrograms and scalogram images. If necessary, perform data augmentation.
2. Define a range for the hyperparameter values of each individual CNN and SAE.
3. Generate $n_1$ random CNNs with hyperparameter values picked up within the preset ranges.
4. Train these CNNs one by one with the spectrogram images.
5. Generate $n_2$ random SAEs with hyperparameter values picked up within the preset ranges.
6. Train these SAEs one by one with the scalogram images.

In the test phase, the following steps are used.

Convert the input sounds into two different images (spectrogram and scalogram images).

1. For each scalogram image, feed the image simultaneously to all SAEs.
2. For each spectrogram image, feed the image simultaneously to all CNNs.
3. In each SAE, extract its predicted label together with the output of the posterior probability corresponding to the scalogram input image.
4. In each CNN, extract its predicted label together with the posterior probability corresponding to the spectrogram input image.
5. Pipe the labels and/or the posterior probability values obtained in steps 4 and 5 into a fusion mechanism to make a final decision on the label of the input.

### C. Network Architecture

To investigate experimentally the proposed method, three different MMDL detectors comprised of 5, 10, and 15 individual CNNs and SAEs, respectively, are generated. For CNNs, each structure is composed of an image input layer, α randomized hidden CNN blocks, a fully connected layer and a Softmax layer. Each randomly generated CNN block is made up of a convolution layer, a batch normalization layer, a ReLU layer, and a max pooling layer. To ensure that the dimension of the captured features decreases after each layer when generating α random numbers (each representing the number of neurons for one CNN layer), we sort these numbers descendingly and then applied them for each of the convolution layers. In the SAE case, one specifies a range in which two hyperparameters, number of AEs in each SAE (*L*), and the number of neurons in each hidden layer are randomly generated. The design of a randomized SAE is done as follows. One first generates *L* random numbers, each representing the number of neurons for one AE. To ensure that the dimension of the captured features is decreasing after each layer, these *L* random numbers are sorted in descending order before being used as the number of hidden neurons of individual AEs. After all AEs are generated, they are stacked together to form an SAE. After all individual CNN and SAE structures are generated, they are trained in parallel. As has been mentioned, SAEs are trained using scalogram images, and CNNs are trained using spectrogram images. Finally, we fuse these results using various methods described in the next subsection.

### D. Fusion Strategies

The fusion of the results obtained by the generated individual models can be done by the following strategies: Majority Vote, Weighted Average, Feed Forward Network (FFN), and a PatternNet. All of the fusion methods conduct fusion operations at the score level with the exception of Majority Voting, which does the task at the decision level. Majority Voting, one of the most popular methods of ensemble, is done by taking the decisions from each of the models and selecting the final decision that has the most votes. Weighted average, another popular fusion method, takes the average of the final fully-connected layer's activations of each CNN model and pipes them into a softmax layer. This method has the advantage over Majority Voting in that it takes into account each model's uncertainty rather than the binary outputs in the decision level. Ju, Bibaut, and van der Laan (Ju et al., 2018) showed that weighted average has the advantage when models have similar performances. Thus, because we intend to take advantage of our model's diversity, we found it appropriate to use another approach.

Our solution strategy of using a PatternNet for model fusion is not to simply throw out less optimal models, nor is it to weigh each model similarly. In our experimental study, we piped the concatenated activations of each individual model into a vector that was then piped into a PatternNet. The PatternNet is essentially an optimized FFN for pattern recognition (Wang et al., 2018). In our study, the cross-entropy measure was adopted as the loss function and the scaled conjugate gradient method as the training procedure. We found that this approach was appropriate as the vector of activations, and in theory, it should hold an overarching periodic pattern due to each model producing slightly similar results. Our MMDL structure was able to learn each individual model's strengths and take advantage of them for final decision making. The PatternNet was trained with a variable number of neurons depending on the number of networks and the number of classes (in our case, we have 2). Number of Neurons is chosen to be $k$ times Number of Classes and times Number of CNN Subnetworks. The constant, $k$, can be heuristically determined by trial and error.

## V. EXPERIMENTAL RESULTS

In order to reduce the impact of commercial shipping on the survival of NARWs, Cornell University researchers deployed an auto-detection buoy network near the Cape Cod Bay area. Cornell University also distributed a dataset of NARW up-calls and non-calls for researchers around the world to devise effective means for NARW up-call detection. One issue with this dataset is that the dataset was initially labeled by using an edge detector, resulting in a substantial number of wrong labels. To mitigate the problem, we converted all the data to spectrogram images and re-labeled the data based on visual inspection of spectrograms and listening of audio signals. In this way, we made sure that our models are being trained with accurate(?) and consistent ground truth data. Our experimental results were divided into two categories. In the first category, handcrafted features with conventional machine learning algorithms were applied for NARW up-call detection. In the second category, the proposed MMDL method was applied to the same problem. Three thousand right whale acoustic segments were used in the training phase, in which there were 241 NARW up-calls and 2,759 non-up-calls. In the verification phase, 1500 NARW acoustic segments were used for the purpose of verification for up-call detection, among which, 233 were considered up-call audio segments, the rest (1,267) were non-up-calls. The following formulas are used to determine detection accuracy:

$$\text{Upcall detection rate} = \frac{T_P}{P}$$

$$\text{Non-upcall detection rate} = \frac{T_N}{N}$$

$$\text{False alarme rate} = \frac{F_P}{N}$$

where True Positive ($T_P$): True is identified True (correct identification); True Negative ($T_N$): False is identified False (correct identification); False Positive ($F_P$): False is identified as True (wrong identification); False Negative ($F_N$): True is identified as False (wrong identification). Thus $P = T_P + F_N$ is total number of up-calls, and $N = T_N + F_P$ is total number of non-up-calls.

It has been found in this research that for classification, Support Vector Machines (SVMs) fares better in comparison to other conventional machine learning algorithms, say K-Nearest Neighborhood (KNN), due to the fact that SVMs perform well on datasets that have overlapped attributes. As a result, the overall experimental system was built by integrating DWTs, MFCCs (or GFCCs), and SVMs.

Table 1 shows the detection results obtained from various types of features and the linear SVM. It is observed that the highest rate of correct up-call detection is achieved by using a 2-stage DWT followed by an MFCC computation, which is 91.85%, and the overall detection rate is 94.47%. On the other hand, the MFCC and SVM combination produces the lowest up-call detection rate.

Table 1. Detection results using MFCCs and DWTs

| Type of feature | TP | FN | FP | FN | Up-call detection | Non-up-call detection | False alarm |
|---|---|---|---|---|---|---|---|
| MFCC | 172 | 1231 | 61 | 36 | 73.82% | 97.86% | 2.48% |
| DWT single stage +MFCC | 181 | 1233 | 52 | 34 | 77.68% | 97.32% | 2.68% |
| DWT two stage +MFCC | 214 | 1203 | 19 | 64 | 91.85% | 94.95% | 5.05% |
| DWT three stage +MFCC | 201 | 1165 | 32 | 103 | 86.27% | 91.96% | 8.05% |

In Table 2, results from using different types of wavelet families (for instance, Symlets and Daubechies) are shown. Table 2 indicates that Daubechies (db4) is the best type of feature for up-call detection, which achieves an accuracy of 92.27%. On the other hand, db1 has the lowest detection rate, which is 86.2661%.

Table 2. Detection results using different wavelets

| Type of Wavelet | TP | TN | FP | FN | Up-call detection | Non-up-call detection | False alarm |
|---|---|---|---|---|---|---|---|
| Db1 | 201 | 1162 | 32 | 105 | 86.27% | 91.69% | 2.68% |
| Db2 | 202 | 1194 | 31 | 73 | 86.70% | 94.24% | 2.53% |
| Db3 | 211 | 1204 | 22 | 63 | 90.56% | 95.03% | 1.79% |
| Db4 | 215 | 1202 | 18 | 65 | 92.27% | 94.87% | 1.48% |
| Db5 | 206 | 1187 | 27 | 80 | 88.41% | 93.69% | 2.22% |
| Db6 | 210 | 1208 | 23 | 59 | 90.13% | 95.35% | 1.87% |
| Db7 | 213 | 1209 | 20 | 58 | 91.41% | 95.43% | 1.63% |
| Coif1 | 207 | 1205 | 26 | 62 | 88.84% | 95.11% | 2.11% |
| Coif2 | 209 | 1191 | 24 | 76 | 89.70% | 94.01% | 1.98% |
| Coif3 | 207 | 1200 | 26 | 67 | 88.84% | 94.72% | 2.12% |
| Coif4 | 212 | 1191 | 21 | 76 | 90.99% | 94.01% | 1.73% |
| Coif5 | 210 | 1204 | 23 | 63 | 90.13% | 95.03% | 1.87% |
| Sym2 | 202 | 1194 | 31 | 73 | 86.70% | 94.24% | 2.53% |
| Sym3 | 211 | 1204 | 22 | 63 | 90.56% | 95.03% | 1.79% |
| Sym4 | 208 | 1191 | 25 | 76 | 89.27% | 94.01% | 2.06% |
| Sym5 | 209 | 1194 | 24 | 73 | 89.70% | 94.24% | 1.97% |
| Sym5 | 201 | 1197 | 32 | 70 | 86.27% | 94.48% | 2.60% |

Table 3. Detection results using different features and different detectors

| Method | TP | TN | FP | FN | Up-call detection | Non-up-call detection | False alarm |
|---|---|---|---|---|---|---|---|
| DWT+MFCC+Linear SVM | 215 | 1202 | 18 | 65 | 92.27% | 94.87% | 1.48% |
| TFP-2 features+LDA | 187 | 1206 | 46 | 61 | 77.68% | 97.32% | 4.81% |
| TFP-2 features+QDA | 157 | 1176 | 76 | 91 | 91.85% | 94.95% | 7.18% |
| TFP-2 features+KNN | 147 | 1213 | 54 | 86 | 63.23% | 95.74% | 4.26% |
| TFP-2 features+Decision Tree | 73 | 1248 | 160 | 19 | 31.18% | 98.52% | 11.36% |
| TFP-2 features+Linear SVM | 165 | 1230 | 68 | 37 | 70.81% | 97.08% | 2.92% |
| TFP-2 features+TreeBagger | 178 | 1230 | 55 | 37 | 76.25% | 97.08% | 2.92% |
| LBP features+LDA | 170 | 1239 | 63 | 28 | 72.96% | 97.82% | 2.21% |
| LBP features+QDA | 183 | 1197 | 50 | 70 | 78.40% | 94.44% | 5.52% |
| LBP features+KNN | 184 | 1208 | 49 | 59 | 78.97% | 95.35% | 4.66% |
| LBP features+Decision Tree | 135 | 1149 | 88 | 118 | 57.79% | 90.65% | 9.31% |
| LBP features+Linear SVM | 211 | 1184 | 22 | 83 | 90.41% | 93.44% | 6.55% |
| LBP features+TreeBagger | 210 | 1184 | 23 | 83 | 89.98% | 93.48% | 6.55% |
| DWT+MFCC+KNN | 144 | 1244 | 89 | 23 | 62.00% | 98.18% | 1.82% |
| MFCC+Linear SVM | 172 | 1231 | 61 | 36 | 73.82% | 97.16% | 2.84% |
| DWT+MFCC+LinearSVM | 215 | 1202 | 18 | 65 | 92.27% | 94.87% | 1.48% |
| DWT+MFCC+KernalSVM (poly) | 171 | 1237 | 62 | 30 | 73.39% | 97.63% | 2.37% |
| DWT+MFCC+Kernal SVM (rbf-100) | 210 | 1166 | 23 | 101 | 90.13% | 92.03% | 7.97% |

Table 3 compares the results obtained in this research project with those reported in (Esfahanian et al., 2015; Motlıcek, 2002). The highest up-call detection rate obtained by using the DWTs & MFCCs as features and the linear SVM as the detector is 92.27% with a low false alarm rate of 1.48%. In comparison, the best up-call detection rate reported in (Esfahanian et al., 2015), which was obtained by using the LBP features and the linear SVMs, is 90.41% with a false alarm rate of 6.55%.

The experimental methodology by using the MMDL detector involved testing three different fusion methods on our proposed MMDL structure with varying numbers of CNN and SAE subnetworks. These networks were tested using 80,665 test files compared to the 1,500 files in the previous results. All results were compiled through a 5-fold cross validation test. Table 4 shows the results of the three representative fusion methods, each having 15 CNN subnetworks on the spectrogram images, and 15 SAE subnetworks on the scalogram images.

Table 4. Results of different ensemble methods (each have 15 CNN models and 5 SAE models)

| Ensemble method | Up-call detection | Non-up-call detection | False alarm |
|---|---|---|---|
| Majority Voting | 97% | 98.7% | 0.23% |
| Unweighted average | 98.1% | 99.08% | 0.12% |
| FFN | 99.8% | 100% | 0.09% |
| PatternNet | 100% | 100% | 0% |

Table 5 shows the individual and ($n$) ensemble results on whole test data, where $n$ indicates the number of randomly generated CNN models in the ensemble architecture. Note that the (overall) results reflect those of the PatternNet ensemble method.

Table 5. Results

| Number of Models | Up-call detection | Non-up-call detection | False alarm |
|---|---|---|---|
| 5 CNNs, 5 SAEs | 99.3% | 99.9% | 0.07% |
| 10 CNNs, 10 SAEs | 99.8 | 100% | 0% |
| 15 CNNs, 15 SAEs | 100% | 100% | 0% |

of different ensemble methods, each employing 15 CNN models,15 SAE models

The results given in Tables 4 and 5 show that increasing the number of models leads to a continued improvement of the classification performance.

## VI. CONCLUSION

In this study, a new approach for NARW up-call detection has been proposed. In this approach, signals from passive acoustic sensors are first converted to spectrogram and scalogram images, and the MMDL detector is then applied for the classification of these images. This deep learning detector is composed of a number of CNNs and SAEs, each of which is randomly designed. Recall CNNs extract discriminative features at both local and global levels, and SAEs are designed for data abstraction and reproduction. Due to the randomness of the model structure and distinct characteristics of CNNs and SAEs, the integrated MMDL detector is more robust against data variability. To validate the effectiveness of the proposed MMDL model for NARW up-call detetction, the dataset published by the Cornell University was used. The Cornel dataset had some mislabeled samples. To correct wrongly labeled data points, the dataset was relabeled by human visual inspection of signal spectrograms and listening of audio signals. With the relabeled dataset, our experimental study demonstrated that the MMDL detector achieved superior performances over conventional machine learning methods in terms of up-call detection rate, non-up-call detection rate, and false alarm rate. It is our belief that the MMDL system can also act as a general classifier for applications in which multiple classes are involved. In comparison to deep networks, an advantage of the MMDL detector is its small footprint, because each sub-model in the system is a shallow network; therefore it is efficient both in terms of its memory requirement and computational complexity.

**ACKNOWLEDGMENTS**

The authors acknowledge the support of the Protect Florida Whales Sepcialty License Plate provided through the Harbor Branch Oceanographic Institute Foundation, and continuing support from Harbor Branch Oceanogrpahic Institute and Florida Atlatic Univerity.. The authors also gratefully acknowledge the acquisition of the acoustic data made available by Cornell University.

**REFERENCES**

Clark, C. W., M. W. Brown, and P. Corkeron. 2010. Visual and acoustic surveys for North Atlantic right whales, Eubalaena glacialis, in Cape Cod Bay, Massachusetts, 2001-2005: Management implications. Mar. Mamm. Sci. 26:837–854.

Clark, C. W. (1982). "The acoustic repertoire of the southern right whale, a quantitative analysis," Animal Behaviour 30(4), 1060-1071.

Cooke, J. (2018). "Eubalaena glacialis. iucn red list of threatened species 2018: e.t41712a50380891".

Hayes SA, Josephson E, Maze-Foley K, and Rosel PE. 2019. US Atlantic and Gulf of Mexico Marine Mammal Stock Assessments – 2018. NOAA Tech. Memo NMFS-NE-258.

De Veth, J., Cranen, B., and Boves, L. (2001a). \Acoustic features and distance measure to reduce vulnerability of asr performance due to the presence of a communication channel and/or back- ground noise," in Robustness in language and speech technology (Springer), pp. 9-45.

Dugan, P. J., Rice, A. N., Urazghildiiev, I. R., and Clark, C. W. (2010). "North atlantic right whale acoustic signal processing: Part i. comparison of machine learning recognition algorithms," in Applications and Technology Conference (LISAT), 2010 Long Island Systems, IEEE, pp. 1-6.

Esfahanian, M., Zhuang, H., Erdol, N., and Gerstein, E. (2015). "Comparison of two methods for detection of north atlantic right whale upcalls," in Signal Processing Conference (EUSIPCO), 2015 23rd European, IEEE, pp. 559-563.

Gillespie, D. (2004). "Detection and classifcation of right whale calls using an'edge'detector operating on a smoothed spectrogram," Canadian Acoustics 32(2), 39-47.

Gu, J., Wang, Z., Kuen, J., Ma, L., Shahroudy, A., Shuai, B., Liu, T., Wang, X., Wang, G., Cai, J. et al. (2018). "Recent advances in convolutional neural networks," Pattern Recognition 77, 354-377.

Hayes, S. A., Josephson, E., Maze-Foley, K., Rosel, P. E., Byrd, B., and Cole, T. (2017). US Atlantic and Gulf of Mexico marine mammal stock assessments-2016 (US Department of Commerce, National Oceanic and Atmospheric Administration ).

Ju, C., Bibaut, A., and van der Laan, M. (2018). "The relative performance of ensemble methods with deep convolutional neural networks for image classification," Journal of Applied Statistics 45(15), 2800-2818.

Krizhevsky, A., Sutskever, I., and Hinton, G. E. (2012). "Imagenet classi_cation with deep convolutional neural networks," in Advances in neural information processing systems, pp. 1097-1105.

Logan, B. et al. (2000). \Mel frequency cepstral coeffcients for music modeling.," in ISMIR, Vol. 270, pp. 1-11.

Mellinger, D. K. (2004). "A comparison of methods for detecting right whale calls," Canadian Acoustics 32(2), 55-65.

Mellinger, D. K., and Clark, C. W. (1993). "A method for _lter-ing bioacoustic transients by spectrogram image convolution," in OCEANS'93. Engineering in Harmony with Ocean. Proceedings, IEEE, pp. III122-III127.

Mermelstein, P. (1976). "Distance measures for speech recognition, psychological and instrumental," Pattern recognition and artificial intelligence 116, 374-388.

Moore, B. C. (2012). An introduction to the psychology of hearing (Brill).

Motl_cek, P. (2002). "Feature extraction in speech coding and recognition," Technical Report .

Oppenheim, A. V., and Schafer, R. W. (2014). Discrete-time signal processing (Pearson Education).

Pace, F., Benard, F., Glotin, H., Adam, O., and White, P. (2010). "Subunit de_nition and analysis for humpback whale call classification," Applied Acoustics 71(11), 1107-1112.

Patterson, R., Nimmo-Smith, I., Holdsworth, J., and Rice, P. (1987). "An efficient auditory _lterbank based on the gammatone function," in a meeting of the IOC Speech Group on Auditory Modelling at RSRE, Vol. 2.

Pylypenko, K. (2015). "Right whale detection using artificial neural network and principal component analysis," in Electronics and Nanotechnology (ELNANO), 2015 IEEE 35th International Conference on, IEEE, pp. 370{373.

Reeves, R. R. (2003). Dolphins, whales and porpoises: 2002-2010 conservation action plan for the world's cetaceans, 58 (IUCN).

Roch, M. A., Soldevilla, M. S., Burtenshaw, J. C., Henderson, E. E., and Hildebrand, J. A. (2007). "Gaussian mixture model classification of odontocetes in the southern california bight and the gulf of california," The Journal of the Acoustical Society of America 121(3), 1737-1748.

Shao, Y., Srinivasan, S., and Wang, D. (2007). "Incorporating auditory feature uncertainties in robust speaker identifcation," in Acoustics, Speech and Signal Processing, 2007. ICASSP 2007. IEEE International Conference on, IEEE, Vol. 4, pp. IV-277.

Urazghildiiev, I. R., and Clark, C. W. (2006). "Acoustic detection of north atlantic right whale contact calls using the generalized likelihood ratio test," The Journal of the Acoustical Society of America 120(4), 1956-1963.

Urazghildiiev, I. R., Clark, C. W., Krein, T. P., and Parks, S. E. (2009). "Detection and recognition of north atlantic right whale contact calls in the presence of ambient noise," IEEE Journal of Oceanic Engineering 34(3), 358-368.

Wang, D., and Brown, G. J. (2006). Computational auditory scene analysis: Principles, algorithms, and applications (Wiley-IEEE press).
24

Wang, J. L., Li, A. Y., Huang, M., Ibrahim, A. K., Zhuang, H., and Ali, A. M. (2018). "Classification of white blood cells with patternnet-fused ensemble of convolutional neural networks (pecnn)," in 2018 IEEE International Symposium on Signal Processing and Information Technology (ISSPIT), IEEE, pp. 325-330.

Young, S., Evermann, G., Gales, M., Hain, T., Kershaw, D., Liu, X., Moore, G., Odell, J., Ollason, D., Povey, D. et al. (2002)."The htk book," Cambridge university engineering department 3, 175.